\input{aipcheck}

\documentclass[
    ,final            
  ]
  {aipproc}

\layoutstyle{8x11single}

\begin{document}

\title{Nonequilibrium liquid theory for sheared granular liquids}


\author{Hisao Hayakawa}{
  address={Yukawa Institute for Theoretical Physics, Kyoto University, Kyoto 606-8502, Japan}
}

\author{Song-Ho Chong}{
  address={Institute for Molecular Science, Okazaki 444-8585, Japan}
}

\author{Michio Otsuki}{
  address={Department of Physics and Mathematics, Aoyama Gakuin University, Sagamihara 229-8558, Japan}
}

\begin{abstract}
A nonequilibrium liquid theory for uniformly sheared granular liquids is developed starting from SLLOD equations.
We derive a generalized Green-Kubo formula and demonstrate that 
it yields the nonequilibrium steady-state average which is essentially independent of the choice of
the initial condition.
It is also shown that the fluctuating hydrodynamics can be derived from Mori-type equations for density and
current-density fluctuations 
if one considers a weak-shear and small-dissipation limit along with the Markovian approximation. 
\end{abstract}

\maketitle


\section{Introduction}

Assemblies of dense granular materials behave unlike usual materials. 
One may think that a liquid phase is absent for granular assemblies since they
lack attractive interaction.
Indeed, the gas kinetic theory such as the Boltzmann-Enskog theory 
has historically been employed in describing dense granular flows, 
where correlation effects appear only 
through the contact value of the radial distribution function.\cite{goldhirsch,brilliantov04,jenkins,garzo,lutsko05,saitoh}
This approach is powerful and semi-quantitatively accurate even for considerably dense systems.
In these days, however, we have recognized the relevancy of the concept of ``granular liquids" because 
correlation effects in granular flows, 
such as long-time correlations and long-range correlations,
turned out to be relevant as in molecular liquids.\cite{kumaran06,orpe07,kudrolli09,kumaran09,otsuki07,otsuki09a,otsuki09b}
 
There have been some developments in the granular liquid theory starting from microscopic basic equations such as
the Liouville equation. 
The Liouville equation of granular fluids was first discussed by Schofield and Oppenheim\cite{oppenheim} long time ago, 
and  Brey {\it et al.}\cite{brey} further developed such a formulation.
Recently, Dufty et al.\cite{dufty08,dufty08b} discussed in detail the Green-Kubo formula of freely cooling granular gases
starting from the Liouville equation. 
The Green-Kubo formula for granular fluids has been discussed  in various
contexts\cite{vanNoije,dufty02a,dufty02b,lutsko02,brey05}, 
and these studies suggest that some correction terms to the conventional Green-Kubo
formula are necessary for granular fluids.
 
There is an advantage in using the liquid theory. 
It is known that liquid theories such as the mode-coupling theory (MCT)
 for supercooled liquids are commonly used to describe the glass transition of molecular liquids 
or colloidal assemblies.\cite{das,goetz}
Liu and Nagel\cite{Liu} proposed that the jamming transition is a fundamental transition in glassy and granular materials. 
Since then many aspects of similarities between the conventional glass transition and the jamming transition have been
investigated \cite{miguel}, where
some researchers  have used granular materials to study dynamical heterogeneity 
in glassy materials. \cite{Daushot05,Abate06,Abate07,Daushot08,Watanabe}
Along this line it appears natural to apply a liquid theory to dense granular assemblies.
 
Nevertheless, the jamming transition under a plane shear exhibits some distinct aspects 
from conventional glass transitions, e.g., 
the jamming transition depends strongly on details of microscopic interactions between particles.
In fact, 
one cannot use a naive MCT~\cite{Hayakawa} in describing sheared jamming transitions.
Instead, the jamming transition of frictionless granular particles is believed to be 
a continuous transition  at a critical density above which elastic moduli and the yield stress become nonzero, and there are scaling laws
in the vicinity of the critical point as observed in 
conventional critical phenomena. \cite{OHern02,OHern03,Olsson,Hatano08_1,Otsuki08a,otsuki09c}

Quite recently,
Chong and Kim\cite{chong} have reformulated the liquid theory of sheared dense molecular liquids with the Gaussian thermostat.
Later, Chong et al.\cite{chong09} extended their formulation to soft granular liquids under a plane shear, and
found the existence of the generalized Green-Kubo formula and integral fluctuation theorem.
Chong et al.\cite{Hayakawa09} have further developed an MCT for uniformly sheared granular liquids.
This generalized Green-Kubo formula applies not only to linear regime but also to 
nonequilibrium states arbitrarily far from equilibrium.
Thus, 
one can avoid the use of hydrodynamic equations including Burnett and super-Burnett terms 
which occasionally exhibit a divergent behavior.

The aim of this paper is to outline our formulation of the liquid theory for dense sheared granular materials.
In the next section, we summarize the microscopic starting equations, such as SLLOD equations, Liouville equations, and
nonequilibrium distribution function.
In section III, we present 
the transient time correlation function formalism and discuss steady state properties. We then derive the generalized Green-Kubo formula.
It is demonstrated that the generalized Green-Kubo formula yields the nonequilibrium steady-state average
which is essentially independent of the specific choice of the initial condition. 
In section IV, 
we outline the liquid theory beyond the generalized Green-Kubo formula,
and discuss its connection to the fluctuating hydrodynamics.
The paper is summarized in section V. 

\section{Microscopic starting equations}

In this section, we derive exact microscopic equations and relations which serve a basis in constructing
a nonequilibrium liquid theory for uniformly sheared frictionless granular particles.

\subsection{SLLOD equations of motion}

Let us consider a system of $N$ smooth granular particles of mass $m$ 
in a volume $V$ under a stationary shear characterized by the
shear-rate tensor $\sf{\kappa}$.
We assume that each granular particle is a soft-sphere, and the contact force acts only on the normal direction.
Under  a homogeneous shear,  
the velocity profile is given by 
$\sf{\kappa} \cdot {\bf r}$
at position ${\bf r}$.
The Newtonian equations of motion describing such a homogeneously 
sheared system are given by the SLLOD equations~\cite{Evans90}
\begin{equation}
\label{eq:SLLOD}
\dot{{\bf r}}_i 
=
\frac{{\bf p}_i}{m} +
\mbox{\boldmath $\kappa$} \cdot {\bf r}_{i},
\end{equation}
where ${\bf r}_{i}$ refers to the position 
of the $i$th particle, $\dot{\bf r}_i=d{\bf r}_i/dt$, and 
\begin{equation}
\dot{{\bf p}}_{i} 
={\bf F}^{\rm (el)}_{i}+{\bf F}^{\rm (vis)}_i - \mbox{\boldmath $\kappa$} \cdot {\bf p}_{i} . 
\label{eq:SLLOD-b}
\end{equation}
Here Eq.(\ref{eq:SLLOD}) is the definition of the peculiar momentum ${\bf p}_i$ satisfying $\sum_i {\bf p}_i=0$,  
${\bf F}^{\rm (el)}_{i} =\sum_{k\ne i} {\bf F}_{ik}^{\rm (el)}$ is the conservative force exerted on the $i$th particle by 
other particles with
\begin{equation}
{\bf F}_{ik}^{\rm (el)}=-\frac{\partial u(r_{ik})}{\partial {\bf r}_{ik}}=\Theta(\sigma-r_{ik})f(d-r_{ik})\hat{{\bf r}}_{ik} ,
\end{equation}
where $\sigma$ is the diameter of each grain, $u(r_{ik})$ is the pairwise potential, ${\bf r}_{ik}={\bf r}_i-{\bf r}_k$, 
$r_{ik}=|{\bf r}_{ik}|$, $\hat{{\bf r}}_{ik}={\bf r}_{ik}/r_{ik}$, and 
$\Theta(x)$ is the Heviside function satisfying $\Theta(x)=1$ for $x>0$ and $\Theta(x)=0$ for $x< 0$.
 The actual elastic repulsive force $f(x)$ is proportional to $x^{3/2}$ for  three dimensional systems,
 but we sometimes use a simpler form $f(x)\propto x$.  
Similarly, the viscous dissipative force ${\bf F}^{\rm (vis)}_i$ is represented by a sum of two-body contact forces
as ${\bf F}^{\rm (vis)}_i=\sum_{j\ne i}{\bf F}^{\rm (vis)}_{ij}$ with
\begin{equation}\label{viscous-force}
{\bf F}^{\rm (vis)}_{ij}=
-\hat{{\bf r}}_{ij}{\cal F}(r_{ij})({\bf g}_{ij}\cdot\hat{{\bf r}}_{ij})
\equiv
-\hat{{\bf r}}_{ij}\Theta(d-r_{ij})
\gamma(\sigma-{r}_{ij})
({\bf g}_{ij}\cdot\hat{{\bf r}}_{ij})
.
\end{equation} 
In Eq. (\ref{viscous-force}) we have introduced 
${\bf g}_{ij}\equiv {\bf v}_i-{\bf v}_j
=({\bf p}_i-{\bf p}_j)/m+\sf{\kappa} \cdot ({\bf r}_i-{\bf r}_j)$ with the velocity of $i$th particle ${\bf v}_i\equiv d{\bf r}_i/dt$.
The viscous function $\gamma(x)$ is proportional to $\sqrt{x}$ for three dimensional systems,  but
we sometimes use a simpler model $\gamma(x)=const$.

It should be noted that SLLOD equations (\ref{eq:SLLOD}) and (\ref{eq:SLLOD-b}) reduce to the Newtonian equation of motion
\begin{equation}
m \ddot{\bf r}_i={\bf F}^{\rm (el)}_{i}+{\bf F}^{\rm (vis)}_i ,
\end{equation}
if one eliminates the peculiar momentum ${\bf p}_i$.
This means that the SLLOD equations are equivalent to 
the Newtonian equation of motion under the Lees-Edwards boundary condition.~\cite{Evans90}
We also note that frictionless sheared granular particles under a constant pressure 
boundary behave  as those under the Lees-Edwards boundary condition in the vicinity
the jamming transition.\cite{HOS} On the other hand, it is hard to extract physical essences from
actual frictional granular assemblies under a physical boundary condition. 
Thus, a set of SLLOD equations is a natural starting point in constructing a liquid theory of granular particles.
We also notice that a granular flow on an inclined slope can be approximately described by a uniformly sheared flow
except for the boundary layers.\cite{Mitarai05}

We stress that  an energy sink term is necessary even for a system of a sheared molecular liquid. Indeed, the system  heats up 
without the energy sink. In simulations, one usually introduces a thermostat, and an experimental apparatus for the real system
plays a role of the thermostat. We have already confirmed that behaviors of uniformly sheared granular liquids with small inelasticity
are almost the same as those for a model of a molecular liquid with a velocity rescaling thermostat.\cite{otsuki07,otsuki09a}    

We have assumed that the interaction between granular particles is described by a soft-core model, which differs from most of 
the conventional treatment of the granular gas kinetic theory and the MCT for sheared granular liquids
where a hard-core model is adopted.\cite{Hayakawa}
The soft-core model has several advantages: (i) the model is more realistic than the hard-core model, 
(ii) one can apply this formulation to very
dense systems even in the vicinity of the jamming transition, and (iii) one can avoid the use of the pseudo-Liouvillian.

\subsection{The Liouville equation}

For nonequilibrium systems described by the SLLOD equations, 
 the Liouville equation is commonly used.~\cite{Evans90}
The time evolution of phase variables
whose time dependence comes solely from that of the phase space point
$\Gamma = ({\bf r}^{N}, {\bf p}^{N})$
is determined by
\begin{equation}
\frac{d}{dt} A(\Gamma) =
\dot{\Gamma} \cdot 
\frac{\partial}{\partial \Gamma}
A(\Gamma) \equiv
i {\cal L} A(\Gamma).
\label{eq:Lp}
\end{equation}
The operator $i{\cal L}$ is referred to as the
Liouvillian.
The formal solution to this equation can be written as
\begin{equation}
A(\Gamma,t) =
\exp(i {\cal L} t)
A(\Gamma).
\label{eq:p-propagator}
\end{equation}

On the other hand, the Liouville equation for the nonequilibrium 
phase-space distribution function $\rho(\Gamma,t)$
is given by
\begin{equation}
\frac{\partial \rho(\Gamma,t)}{\partial t} =
- \Bigl[ \, \dot{\Gamma} \cdot 
       \frac{\partial}{\partial \Gamma} +
       \Lambda(\Gamma) \,
\Bigr] \rho(\Gamma,t) \equiv
- i {\cal L}^{\dagger} \rho(\Gamma,t),
\label{eq:Lf}
\end{equation}
where the phase space contraction factor $\Lambda(\Gamma)$ 
is defined by
\begin{equation}
\Lambda(\Gamma) \equiv
\frac{\partial}{\partial \Gamma} \cdot
\dot{\Gamma}
=\sum_{i}
\Bigl( \,
  \frac{\partial}{\partial {\bf r}_{i}} \cdot \dot{{\bf r}}_{i} +
  \frac{\partial}{\partial {\bf p}_{i}} \cdot \dot{{\bf p}}_{i} \,
\Bigr)  .
\label{eq:Lambda}
\end{equation}
For our model (\ref{eq:SLLOD}) - (\ref{viscous-force})
one easily obtains its explicit form: 
\begin{equation}
\Lambda(\Gamma)=-\frac{1}{m}\sum_{i,j}\Theta(d-r_{ij})\gamma(d-r_{ij})<0 .
\label{eq:SLLOD-Lambda}
\end{equation}
The formal solution to the Liouville equation (\ref{eq:Lf}) reads
\begin{equation}
\rho(\Gamma,t) = \exp( - i {\cal L}^{\dagger} t) \,
\rho(\Gamma,0).
\label{eq:f-propagator}
\end{equation}
From Eqs.~(\ref{eq:Lp}) and (\ref{eq:Lf}) we readily obtain the relation
\begin{equation}
i{\cal L}^{\dagger}(\Gamma) = 
i{\cal L}(\Gamma) + \Lambda(\Gamma).
\label{eq:relation-Liouville-operators}
\end{equation}
One can show that
the following adjoint relations hold~\cite{Evans90}:
\begin{eqnarray}\label{eq:adjoint-property}
\int d\Gamma \,
[ i {\cal L} A(\Gamma) ] \, B(\Gamma) &=&
- \int d\Gamma \,
A(\Gamma) \,
[ i {\cal L}^{\dagger} B(\Gamma)], \\
\int d\Gamma [e^{i{\cal L} t} A(\Gamma))]B(\Gamma) &=& \int d\Gamma A(\Gamma) e^{-i {\cal L}^{\dagger} t}B(\Gamma) .
\label{eq:adjoint-property2}
\end{eqnarray}

If the phase-space contraction factor $\Lambda(\Gamma)$
is identically zero, then $i{\cal L}^{\dagger} = i{\cal L}$ holds, 
and the Liouvillian becomes self-adjoint or Hermitian.
In general, this is not the case for nonequilibrium dissipative systems.

\subsection{Nonequilibrium distribution function}

Let us consider an equilibrium system
to which a constant shear rate $\dot{\gamma}$ satisfying $\kappa_{\alpha\beta}=\dot\gamma \delta_{\alpha x}\delta_{\beta y}$ is applied
at time $t=0$, and thereafter the system evolves according to
the SLLOD equations (\ref{eq:SLLOD}) and (\ref{eq:SLLOD-b}).
The Liouvillian is given by $i{\cal L}= i {\cal L}^{\rm (el)}$ for $t<0$ and
\begin{equation}
i {\cal L} = 
i {\cal L}^{\rm (el)} + i {\cal L}_{\dot{\gamma}} + i {\cal L}^{\rm (vis)} 
\label{eq:iL-SLLOD-a}
\end{equation}
for $t>0$, 
where
the unperturbed adiabatic or the elastic part ($i {\cal L}^{\rm (el)}$),
the shear part $(i {\cal L}_{\dot{\gamma}})$,
and the viscous part $(i {\cal L}^{\rm (vis)} )$ 
are respectively given by
\begin{eqnarray}
i {\cal L}^{\rm (el)} &=& \sum_{i}
\Bigl[ \,
  \frac{{\bf p}_{i}}{m} \cdot \frac{\partial}{\partial {\bf r}_{i}} +
  {\bf F}_{i}^{\rm (el)} \cdot \frac{\partial}{\partial {\bf p}_{i}} \,
\Bigr],
\label{eq:iL0}
\\
i {\cal L}_{\dot{\gamma}} &=& \sum_{i}
\Bigl[ \,
  (\sf{\kappa} \cdot {\bf r}_{i}) \cdot  
  \frac{\partial}{\partial {\bf r}_{i}} -
  (\sf{\kappa} \cdot {\bf p}_{i}) \cdot  
  \frac{\partial}{\partial {\bf p}_{i}} \,
\Bigr],
\label{eq:iL-dot-gamma}
\\
i {\cal L}^{\rm (vis)} &=& \sum_{i}
  {\bf F}_i^{\rm (vis)} \cdot 
\frac{\partial}{\partial {\bf p}_{i}}.
\label{eq:iL-vis}
\end{eqnarray}
Here, we assume that the initial distribution is given by the canonical one
\begin{equation}\label{canonical-ini}
\rho(\Gamma,0)=\rho_{\rm eq}(\Gamma)\equiv \frac{e^{-\beta H(\Gamma)}}{Z(\beta)} ; \qquad Z(\beta)\equiv \int d\Gamma e^{-\beta H(\Gamma)},
\end{equation}
where $\beta\equiv 1/T$ is the inverse temperature in the initial state, and  $H$ is the total Hamiltonian defined by
\begin{equation}\label{hamiltonian}
H=\sum_i \left\{ \frac{{\bf p}_i^2}{2m}+\frac{1}{2}\sum_{k\ne i} u(r_{ik}) \right\} .
\end{equation}
It should be noted that the effect of the shear appears in $H$ through 
${\bf p}_i^2/2m=m({\bf v}_i-\sf{\kappa}\cdot{\bf r}_i)^2/2$. We also note that there holds a trivial relationship
\begin{equation}\label{L^eq_rho=0}
i {\cal L}^{\rm (el)} \rho_{\rm eq} =0.
\end{equation}

It might appear that our formulation depends strongly on 
our choice of the initial condition, Eq. (\ref{canonical-ini}). 
However, we will argue below that nonequilibrium steady-state properties therefrom 
are insensitive to such a choice. 

From Eq. (\ref{eq:Lf}) the time evolution of the distribution function can be written as   
\begin{equation}\label{eq:dist-dum-11}
\rho(\Gamma,t)=e^{-i{\cal L}^{\dagger}t}\rho_{\rm eq}(\Gamma) .
\end{equation}
With the identity
\begin{equation}
e^{- i {\cal L}^{\dagger} t} =
1 + \int_{0}^{t} ds \,
e^{- i {\cal L}^{\dagger} s} 
(- i {\cal L}^{\dagger}),
\end{equation}
Eq.~(\ref{eq:dist-dum-11}) can be expressed as
\begin{equation}
\rho(\Gamma,t) = \rho_{\rm eq}(\Gamma) +
\int_{0}^{t} ds \, 
e^{- i {\cal L}^{\dagger} s}
(- i {\cal L}^{\dagger})
\rho_{\rm eq}(\Gamma).
\label{eq:dist-dum-12}
\end{equation}
From Eqs.~(\ref{eq:SLLOD-Lambda}), (\ref{eq:relation-Liouville-operators}), (\ref{eq:iL-SLLOD-a})-(\ref{eq:iL-vis}),
and (\ref{L^eq_rho=0})
we get \begin{equation}
i {\cal L}^{\dagger} \rho_{\rm eq}(\Gamma) =
i {\cal L}_{\dot{\gamma}} \rho_{\rm eq}(\Gamma) +
i {\cal L}^{\rm (vis)} \rho_{\rm eq}(\Gamma) +\Lambda(\Gamma)\rho_{\rm eq}(\Gamma) .
\label{eq:dist-dum-13}
\end{equation}
The first term on the right hand side in this expression is given by
\begin{eqnarray}
i {\cal L}_{\dot{\gamma}} \rho_{\rm eq}(\Gamma) &=&
\beta\sum_{i}
\Bigl[ 
  (\sf{\kappa} \cdot {\bf r}_{i}) \cdot {\bf F}_{i}^{\rm (el)} +  
  (\sf{\kappa} \cdot {\bf p}_{i}) \cdot 
  \frac{{\bf p}_{i}}{m} 
\Bigr] \, \rho_{\rm eq}(\Gamma) \nonumber \\
&=&
\sf{\kappa} : {\sf \sigma}^{\rm (el)}\rho_{\rm eq}(\Gamma) 
=\dot{\gamma}\sigma_{xy}^{\rm (el)}\rho_{\rm eq}(\Gamma)
\label{eq:dist-dum-14}
\end{eqnarray}
where ${\sf \sigma}^{\rm (el)}$ denotes the elastic stress tensor whose
element is given by 
\begin{equation}
\sigma_{\alpha \beta}^{\rm (el)} =
\sum_{i}
[ \, \frac{p_{i}^{\alpha} p_{i}^{\beta} }{ m} + r_i^{\alpha} F_i^{\rm (el) \beta} \, ]
\quad
(\alpha, \beta = x, y, z).
\label{eq:sigma-def}
\end{equation}
In the final equality of Eq.~(\ref{eq:dist-dum-14}) we have used the
specific form $\kappa_{\alpha \beta} = 
\dot{\gamma} \delta_{\alpha x} \delta_{\beta y}$.
On the other hand, from Eq. (\ref{eq:iL-vis}), $i {\cal L}^{\rm (vis)}\rho_{\rm eq}(\Gamma)$ is given by
\begin{eqnarray}
i {\cal L}^{\rm (vis)}\rho_{\rm eq}(\Gamma)&=&
- \beta \rho_{\rm eq}(\Gamma)
\sum_{i} {\bf F}_{i}^{\rm (vis)} \cdot \frac{{\bf p}_{i}}{m}
\nonumber \\
&=&
\beta \Bigl[ \, \sum_{i} (\sf{\kappa} \cdot {\bf r}_{i}) \cdot {\bf F}_{i}^{\rm (vis)} \, \Bigr]
\rho_{\rm eq}(\Gamma)
-\frac{\beta}{2}\sum_{i,k}{\bf F}_{ik}^{\rm (vis)}\cdot{\bf g}_{ik}\rho_{\rm eq}(\Gamma) .
 \label{eigen-vis}
\end{eqnarray}

It is convenient to introduce Rayleigh's dissipation function ${\cal R}$ as
\begin{equation}
{\cal R}\equiv - \frac{1}{4}\sum_{i,k}{\bf g}_{ik}\cdot {\bf F}^{\rm (vis)}_{ik}
=\frac{1}{4}\sum_{i,k}\Theta(d-r_{ik})\gamma(d-r_{ik})({\bf g}_{ik}\cdot\hat{{\bf r}}_{ik})^2.
\label{Rayleigh}
\end{equation}
$\rho(\Gamma,t)$ can then be written as
\begin{equation}\label{rho(gamma)}
\rho(\Gamma,t)=
\rho_{\rm eq}(\Gamma)+
\int_0^tds e^{-i{\cal L}^{\dagger} s}[\rho_{\rm eq}(\Gamma)\Omega(\Gamma)],
\end{equation}
where $\Omega(\Gamma)$ is the nonequilibrium work function defined by
\begin{equation}\label{Omega}
\Omega(\Gamma)\equiv 
- \beta \dot{\gamma} \sigma_{xy}(\Gamma) - 2 \beta
{\cal R}(\Gamma) - \Lambda(\Gamma)
\end{equation}
in terms of the total stress tensor
\begin{equation}
\sigma_{\alpha \beta} =
\sum_{i}
[ \, \frac{p_{i}^{\alpha} p_{i}^{\beta} }{ m} + 
r_i^{\alpha} (F_i^{\rm (el) \beta} + F_i^{\rm (vis) \beta}) \, ]
\quad
(\alpha, \beta = x, y, z).
\end{equation}
One can show that the
equilibrium average of the nonequilibrium work function is zero:
\begin{equation}\label{Omega-properties}
\langle \Omega(\Gamma) \rangle_{\rm c} = 0.
\end{equation}Hereafter, we shall reserve the notation
$\langle \cdots \rangle_{\rm c}$
for representing the averaging over the
initial canonical distribution function
$\rho(\Gamma,0) = \rho_{\rm eq}(\Gamma)$: 
\begin{equation}
\langle \cdots \rangle_{\rm c} \equiv
\int d\Gamma \, \rho_{\rm eq}(\Gamma) \cdots.
\label{eq:def-averaging}
\end{equation}

\section{generalized Green-Kubo formula}

\subsection{Transient time-correlation function formalism and generalized Green-Kubo formula}

In contrast to the case of equilibrium quantities,
the nonequilibrium ensemble average 
$\langle A(t) \rangle_{\rm c}$ 
of a phase variable $A(t)$
depends explicitly on the time $t$ 
past since the start of the shearing.
Using the nonequilibrium phase-space distribution
function $\rho({\Gamma},t)$, 
$\langle A(t) \rangle_{\rm c}$ 
can be expressed as
\begin{eqnarray}
& &
\hspace{-0.5cm}
\langle A(t) \rangle_{\rm c} =
\int d{\Gamma} \,
\rho({\Gamma},0) \, A(t) = 
\int d{\Gamma} \,
\rho({\Gamma},t) \, A(0),
\label{eq:neq-tcf}
\end{eqnarray}
where the second equality follows from Eq.~(\ref{eq:adjoint-property2}). 

Substituting Eq.~(\ref{rho(gamma)}) into 
Eq.~(\ref{eq:neq-tcf})
and then using Eq.~(\ref{eq:adjoint-property}), 
one obtains
\begin{equation}
\langle A(t) \rangle_{\rm c}  = 
\langle A(0) \rangle_{\rm c} 
+\int_0^t ds \langle A(s) \Omega(0) \rangle_{\rm c} .
\label{eq:neq-average-TTCF}
\end{equation}

The expression (\ref{eq:neq-average-TTCF})
relates the nonequilibrium value of a phase variable $A(t)$ at
time $t$ to the integral of 
the {\em transient} time-correlation function. 
Indeed the integrand in Eq. (\ref{eq:neq-average-TTCF}), 
$\langle A(s) \Omega(0) \rangle_{\rm c}$, is  
the correlation between the nonequilibrium work function
in the initial state
and $A$ at time $s$ after the shearing force is 
turned on.
It should be remembered, however, that
the dynamics inside the brackets $\langle \cdots \rangle_{\rm c}$ 
is governed by the
granular SLLOD equations, 
and only averages like 
$\langle A(0) \rangle_{\rm c}$ 
coincide with equilibrium quantities.

The system is  in a nonequilibrium steady state
if the ensemble averages of all phase variables become 
time-independent. 
Let us notice that 
the long-time limit of Eq.~(\ref{eq:neq-average-TTCF})
approaches a constant, and hence, the integral is convergent if the system 
displays {\em mixing}.~\cite{Evans90}
This feature can be demonstrated by taking a time
derivative of Eq.~(\ref{eq:neq-average-TTCF}):
\begin{equation}
\frac{d}{dt} \langle A(t) \rangle_{\rm c} 
= \langle A(t)\Omega(0) \rangle_{\rm c} .
\label{eq:dA-over-dt}
\end{equation}
If the system displays mixing~\cite{Evans90}, all long-time correlations between phase variables vanish.
With the aid of Eq.(\ref{Omega-properties})  we obtain
 $\lim_{t\to \infty}(d/dt)\langle A(t)\rangle_{\rm c} =0$
and
\begin{equation}
\lim_{t \to \infty} \langle A(t) \rangle_{\rm c} =
\langle A \rangle_{\rm ss}.
\end{equation}
Here, the steady-state average, denoted by
$\langle \cdots \rangle_{\rm ss}$ hereafter,
is obtained from
the long-time limit of Eq.~(\ref{eq:neq-average-TTCF}):
\begin{equation}
\langle A \rangle_{\rm ss}  = 
\langle A(0) \rangle_{\rm c} +\int_0^{\infty}
ds \langle A(s)\Omega(0) \rangle_{\rm c} .
\label{eq:ss-average}
\end{equation}
Because of Eq.(\ref{Omega-properties}), Eq.(\ref{eq:ss-average}) can be rewritten as
\begin{equation}
\langle A \rangle_{\rm ss}  = 
\langle A(0) \rangle_{\rm c} +\int_0^{\infty}
ds \langle \Delta A(s)\Omega(0) \rangle_{\rm c} ,
\label{eq:ss-average-2}
\end{equation}
where $\Delta A(s)\equiv A(s)-A(s\to\infty)$. 

Equation~(\ref{eq:ss-average}) is the generalized
Green-Kubo formula which relates the steady-state average to the
time-correlation function describing transient dynamics evolving from 
an initial equilibrium towards a final steady state.
One can easily show that Eq.~(\ref{eq:ss-average})
reduces to the conventional Green-Kubo formula 
if the external force is weak and the dissipative force is neglected,
i.e., for small $\dot{\gamma}$ and $\gamma(x) = 0$. 
For example, by setting $A = \sigma_{xy}$ in
Eq.~(\ref{eq:ss-average}), one obtains
for the steady-state shear stress defined via 
$\sigma_{\rm ss} \equiv - \langle \sigma_{xy} \rangle_{\rm ss} / V$
\begin{equation}
\sigma_{\rm ss} = - \frac{\langle \sigma_{xy}(0) \rangle_{\rm c}}{V} - \frac{1}{V}
\int_{0}^{\infty} ds \,
\langle \sigma_{xy}(s) \Omega(0) \rangle_{\rm c}.
\label{eq:sigma-ss}
\end{equation}
When $\gamma(x) = 0$, 
there hold $\langle \sigma_{xy}(0) \rangle_{c} = 0$ and
$\Omega = - \beta \dot{\gamma} \sigma_{xy}$ [see Eq.~(\ref{Omega})], and
Eq.~(\ref{eq:sigma-ss}) formally reduces to
\begin{equation}
\sigma_{\rm ss} = \frac{\beta \dot{\gamma}}{V} \int_{0}^{\infty} ds \, \langle \sigma_{xy}(s) \sigma_{xy}(0) \rangle_c.
\label{eq:sigma-Green-Kubo}
\end{equation}
For small $\dot{\gamma}$, one can replace the Liouvillian governing the dynamics of
$\sigma_{xy}(s)$ in the integrand by that for a quiescent equilibrium state, and hence,
Eq.~(\ref{eq:sigma-Green-Kubo}) is the conventional Green-Kubo formula 
for the viscosity $\eta$ defined via 
$\eta \equiv \sigma_{\rm ss} / \dot{\gamma}$. 
It should be noted that the conventional derivation of Green-Kubo formula requires a convergent factor  $e^{-\epsilon t}$ 
for the integrand with taking the limit $\epsilon\to 0$.  
Similarly, a small dissipation is also necessary for Eq.(\ref{eq:sigma-ss}) to obtain a convergent result.
Namely, the dissipation plays a role of the convergent factor.
 Otherwise, the system is heated up and cannot reach a steady state.
 
Therefore, with the aid of Eq.(\ref{eq:sigma-ss}), the  viscosity $\eta$
satisfies
\begin{equation}
\eta=- \frac{\langle \sigma_{xy}(0) \rangle_{\rm c}}{\dot\gamma V} - \frac{1}{\dot\gamma V}
\int_{0}^{\infty} ds \,
\langle \sigma_{xy}(s) \Omega(0) \rangle_{\rm c}
\label{viscosity}
\end{equation}
in general situations. 
We should stress that Eq. (\ref{eq:sigma-ss}) or Eq.(\ref{viscosity}) is the full-order expression,
and applies to nonequilibrium states arbitrarily far from equilibrium. 
Thus, we do not have to worry about Burnett or super-Burnett terms 
which occasionally exhibit an unstable behavior. 
In other words, the viscosity $\eta$ or the steady shear stress $\sigma_{\rm ss}$
involves effects of nonlinear rheology, and is free from the magnitude of the deviation from a reference state.  

\subsection{On the initial condition}

In this subsection, let us demonstrate that $\langle A \rangle_{\rm ss}$ is 
independent of the choice of the initial condition such as the initial temperature and initial distribution.\cite{chong09} This result is highly
nontrivial, because Eq. (\ref{eq:ss-average}) appears to depend on the choice of the initial canonical distribution.

From Eqs.~(11) and (19) one obtains 
the  Kawasaki representation~\cite{Evans90}
\begin{equation}
\rho(\Gamma,t)
=
\exp\Bigl[ - \int_{0}^{t} ds \, \Lambda(-s) \Bigr]
\frac{e^{-\beta H(-t)}}{Z(\beta)}
=\rho_{\rm eq}(\Gamma)
\exp\Bigl[ \int_{0}^{t} ds \, \Omega(-s) \Bigr] .
\label{eq:new-neq-distribution}
\end{equation}
In the second equality we have introduced the
nonequilibrium work function $\Omega(t) = e^{i {\cal L} t} \Omega(\Gamma)$ at time $t$.

Since 
\begin{equation}
\frac{\partial}{\partial \beta} 
\Bigl\{ \frac{e^{-\beta H(-t)}}{Z(\beta)} \Bigr\} =
[ \langle H \rangle_{\rm c} - H(-t) ] \, 
\frac{e^{-\beta H(-t)}}{ Z(\beta)},
\label{eq:dA-dbeta-dum-03}
\end{equation}
one obtains from Eqs.~(\ref{eq:new-neq-distribution}) and the definition of the average
for $t \to \infty$
\begin{eqnarray}
\frac{\partial}{\partial \beta} 
\langle A(t) \rangle_{\rm c} &=&
\int d\Gamma \, 
A(0) \, [ \langle H \rangle_{\rm c} - H(-t) ] \, \rho(\Gamma,t)
\nonumber \\
&=&
\langle A(t) \rangle_{\rm c} 
\langle H \rangle_{\rm c} -
\langle A(t) H \rangle_{\rm c} \to 0,
\label{eq:dA-dbeta-dum-04}
\end{eqnarray}
i.e., 
$\langle A \rangle_{\rm ss} = \lim_{t \to \infty} \langle A(t) \rangle_{\rm c}$
is independent of the inverse temperature $\beta$
of the initial equilibrium state.
Thus, 
$\langle A \rangle_{\rm ss}$ 
is uniquely specified by the ``thermodynamic'' parameters
$(N, V, \dot{\gamma})$ characterizing the nonequilibrium steady state.

One can prove a stronger statement that the average of any variable in the steady state is invariant 
if the initial condition can be expanded in an orthogonal polynomial of the kinetic energy associated 
with the Gaussian function or the exponential function
such as the Laguerre bi-polynomial and the Hermite polynomial. Let us demonstrate that the average under the initial condition
expanded in the Sonine polynomial, which is related to the Laguerre bi-polynomial, is the same as the one under the canonical initial condition.
We assume  the following initial condition
\begin{equation}\label{sonine}
\rho_{\rm in}(\Gamma)\equiv
\frac{e^{-\beta H}}{Z(\beta)}\left\{1+\sum_{l=1}^{\infty} a_l S_{3/2}^{(l)}\left(\beta \frac{p^2}{2m}\right)\right\} 
 , \qquad
p^2\equiv \sum_i {\bf p}_i^2 ,
\end{equation}     
where $a_l$ and  $S_{3/2}^{(l)}(x)$ are respectively the expansion parameter and  the Sonine polynomial which satisfies
the orthogonal condition
\begin{equation}\label{vertical_con}
\int_0^{\infty} dx e^{-x} x^m S_m^{(p)}(x)S_m^{(q)}(x)=\frac{\Gamma(m+p+1)}{p!}\delta_{p,q}
\end{equation}
with the Gamma function $\Gamma(x)$. 
It should be noted that the Sonine expansion around the Gaussian has widely been used 
for the description of freely cooling granular gases\cite{brilliantov04}, but Eq.(\ref{sonine}) is more general
than the case of freely cooling cases. Indeed, any function of the kinetic energy 
can be expanded in an orthogonal polynomial. Thus, the only assumption adopted here is that
the initial condition can be represented by a product of the canonical distribution and a function of the kinetic energy. 

Let us denote the average under $\rho_{\rm in}(\Gamma)$ as 
\begin{equation}
\langle A(t) \rangle_{\rm in}\equiv \int d\Gamma \rho_{\rm in}(\Gamma) A(t) . 
\end{equation}
We also introduce the difference between the average under the initial condition (\ref{sonine})
and the one under the canonical initial condition (\ref{canonical-ini}):
\begin{equation}
\delta A(t)\equiv \langle A(t) \rangle_{\rm in}-\langle A(t) \rangle_{\rm c} .
\end{equation}
One immediately obtains
\begin{eqnarray}\label{ini-5}
\delta A(t) &= &\int d\Gamma A(\Gamma)(\rho_{\rm in}(\Gamma(t))-\rho_{\rm c}(\Gamma(t)))  \nonumber\\
&=&  \int d\Gamma A(\Gamma,t)(\rho_{\rm in}(\Gamma)-\rho_{\rm c}(\Gamma)) ,
\end{eqnarray}
where we have used Eq.(\ref{eq:adjoint-property2}). 
Substituting (\ref{sonine}) into (\ref{ini-5})  we obtain
\begin{eqnarray}\label{ini-7}
\delta A(t)&=& 
\sum_{l=1}^{\infty} a_l \int d\Gamma \frac{e^{-\beta H}}{Z(\beta)} A(\Gamma,t)S_{3/2}^{(l)}\left(\beta \frac{p^2}{2m}\right)
\nonumber\\
&=& \sum_{l=1}^{\infty} a_l \left\langle A(\Gamma,t)  S_{3/2}^{(l)}\left(\beta \frac{p^2}{2m}\right) \right\rangle_{\rm c}
\to \sum_{l=1}^{\infty}a_l \langle A(\Gamma,t)\rangle_{\rm c} \left\langle S_{3/2}^{(l)}\left(\beta \frac{p^2}{2m}\right) \right\rangle_{\rm c}
\end{eqnarray}
in the limit $t\to \infty$ with the aid of the mixing property.
If we use 
$S_{3/2}^{(0)}(x)=1$ and 
Eq. (\ref{vertical_con}),  we get the relation
\begin{equation}
\left\langle S_{3/2}^{(l)}\left(\beta \frac{p^2}{2m}\right) \right\rangle_{\rm c}=0 .
\end{equation}
Thus, we obtain
\begin{equation}\label{independent}
\lim_{t\to\infty} \delta A(t)=0 ,
\end{equation}
which is the end of proof.
Thus, the steady state starting from Eq.(\ref{sonine}) is equivalent to the one 
from the canonical distribution (\ref{canonical-ini}).

To demonstrate the irrelevancy of the choice of a specific initial condition, 
we shall present computer-simulation results for two-dimensional soft granular particles.
The simulations have been done for the canonical initial condition and for the homogeneous cooling state.
The system consists of polydisperse 4000 grains of diameters 
$0.7\sigma_0$, $0.8\sigma_0$, $0.9\sigma_0$ and $\sigma_0$, and the number of grains of each diameter is 1000. 
The total area fraction is $\phi=0.5$.
We have adopted the linear spring model to represent the elastic repulsion during a contact.
All variables are non-dimensionalized by the maximum diameter of grains $\sigma_0$, 
its mass $m$ and the spring constant $k$.
The dissipation appears through the linear viscous damping with its coefficient 
$\eta_{\rm damp}=1.0$. 
($m$, $k$, and $\eta_{\rm damp}=1.0$ are common to all the grains.)
The applied shear rate is $\dot \gamma = 0.0005$. 
The freely cooling initial condition has been prepared by performing 
a granular simulation in the absence of shear
up to $t=200$ starting from the canonical distribution at $T_0=5\times 10^{-5}$.

Figure 1 shows the result of the granular temperature defined by the kinetic energy. 
It is easily seen that all the results starting from different initial conditions
converge to a unique steady kinetic temperature in the long time limit.
\begin{figure}
  \includegraphics[height=.3\textheight]{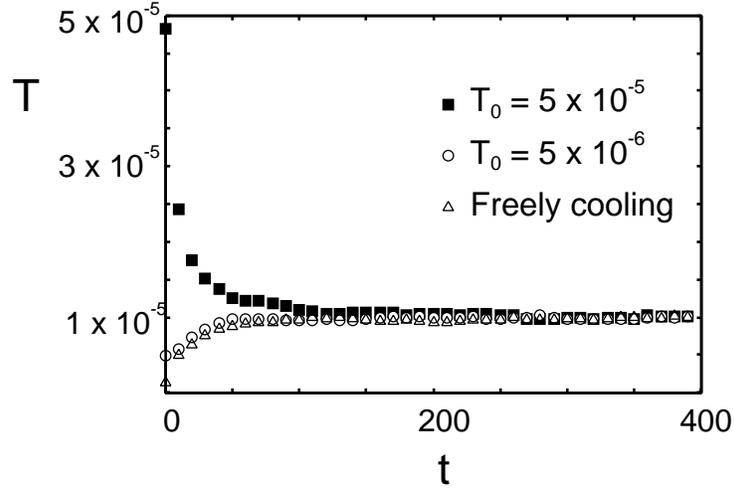}
  \caption{The time evolution of the granular temperature starting from $T_0=5\times 10^6$, $T_0=5\times 10^5$, and
 the homogeneous cooling state. See the text in details.}
\end{figure}

Figure 2 displays the radial distribution function  $g(r)$ for the largest grains in a steady state.
We find that the radial distribution function $g(r)$ in the steady state is independent of the initial condition.
Thus, our numerical results support our theoretical prediction that
steady-state properties are independent of the choice of a specific initial condition.

\begin{figure}
  \includegraphics[height=.3\textheight]{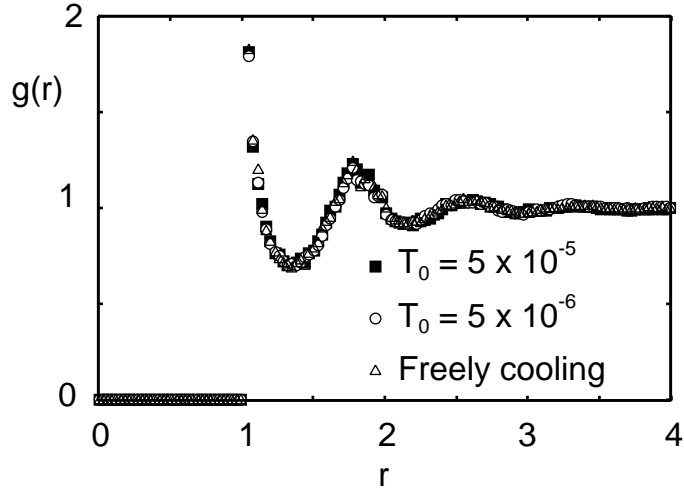}
  \caption{The radial distribution function in the steady state. The legend is common with that in Fig.1.}
\end{figure}

Thus, the generalized Green-Kubo formula (\ref{eq:sigma-ss}) yields the same steady-state average
irrespective of the initial condition.
We should note that  most of nonequilibrium generalizations of the Green-Kubo formula assume the existence of a nonequilibrium
steady distribution function and strongly depends on its steady distribution $\rho_{\rm ss}$.
This standard method has several difficulties such as (i) the determination of $\rho_{\rm ss}$ is difficult, and
(ii) the distribution might not approach a steady value, because there are in general no compatible solutions of both
 $i {\cal L} A_{\rm ss} (\Gamma)=0$ and $i{\cal L}^{\dagger}\rho_{\rm ss}(\Gamma)=0$. 
On the other hand, our formulation is free from such difficulties, and we can calculate,
e.g., the steady shear stress under the canonical initial condition and the obtained result is
independent of the adopted initial condition. 

\section{Outline of the Liquid theory beyond Green-Kubo formula}

In this section, we briefly explain how to use the generalized Green-Kubo formula to describe sheared granular liquids.
Because of the limitation of the length of this paper, we shall skip details of the derivation, and the
interested reader is referred to ref.\cite{Hayakawa09}.

Because the energy is not a conserved quantity which quickly relaxes to a steady value in the uniform shear, 
the relevant hydrodynamic variables are the density fluctuations
\begin{equation}
n_{{\bf q}} (t)\equiv
\sum_{i} e^{i {{\bf q}} \cdot {{\bf r}}_{i}(t)} - N \delta_{{\bf q},{\bf 0}},
\label{eq:rho-def}
\end{equation}
and
the current density fluctuations $j_{{\bf q}}^{\lambda}$ defined by
\begin{equation}
j_{{\bf q}}^{\lambda} = \sum_{i} 
\frac{p_{i}^{\lambda}}{m} 
e^{i {\bf q} \cdot {\bf r}_{i}}.
\label{eq:j-def}
\end{equation}
We introduce
the projection operator ${\cal P}$ onto these variables:
\begin{equation}
{\cal P} X \equiv
\sum_{{\bf k}}
\langle X n_{{\bf k}}^{*} \rangle
\frac{1}{N S_{k}} n_{{\bf k}} +
\sum_{{\bf k}}
\langle X j_{{\bf k}}^{\mu \, *} \rangle
\frac{1}{N v_T^{2}} j_{{\bf k}}^{\mu},
\label{eq:P-tr-def}
\end{equation}
where $S_k\equiv \langle n_{{\bf k}}(t)n_{{\bf k}}(t)^* \rangle/N$ and $v_T\equiv \sqrt{T/m}$.
The complementary projection operator shall be defined by
${\cal Q} \equiv I - {\cal P}$. 

Let us introduce
\begin{equation}
R_{\bf q}^{\lambda}(t) \equiv 
e^{i {\cal QLQ} t} R_{{\bf q}}^{\lambda}
\label{eq:R-tr-def-1}
\end{equation}
with
\begin{eqnarray}
R_{{\bf q}}^{\lambda} &\equiv& {\cal Q} i \tilde{\cal L}
j_{{\bf q}}^{\lambda} 
=
i \tilde{\cal L} j_{{\bf q}}^{\lambda} 
- i q_{\lambda}
\frac{v_T^{2}}{S_{q}} n_{{\bf q}} -
i B_{{\bf q}}^{\lambda}n_{{\bf q}} +
A_{{\bf q}}^{\lambda \mu} j_{{\bf q}}^{\mu},
\label{eq:R-tr-def-2}
\end{eqnarray}
where $i \tilde{{\cal L}}\equiv i {\cal L}^{\rm (el)}+i {\cal L}^{\rm (vis)}$, and
\begin{eqnarray}
A_{{\bf q}}^{\lambda \mu} &=&
\frac{n}{m}
\int d{\bf r} \, 
( 1-e^{i {{\bf q}} \cdot {{\bf r}}} ) \,
\hat{r}^{\lambda} \hat{r}^{\mu} \, {\cal F}(r) g(r),
\label{eq:def-of-A}
\\
iB_{{\bf q}}^{\lambda} &=&
-  \,
\frac{\dot{\gamma}}{mS_q}
\Bigl\{
n \int d{\bf r} \,
\hat{r}^{\lambda} \hat{x} \hat{y} \, r {\cal F}(r) g(r) \,
e^{i {\bf q} \cdot {\bf r}} +
n^{2} \int d{\bf r} \int d{\bf r}' \,
\hat{r}^{\lambda} \hat{x} \hat{y} \, r {\cal F}(r) g^{(3)}({\bf r}, {\bf r}') \,
e^{i {\bf q} \cdot {\bf r}'}
\Bigr\} ,
\label{def-of-AB}
\end{eqnarray}
in terms of the pair and triple correlation functions
\begin{eqnarray}
n g(r) &=& \frac{1}{N} \sum_{i,j \atop i \ne j} \langle \delta({{\bf r}} - 
{\bf r}_{ij}) \rangle,
\label{eq:g2-def}
\\
n^2g^{(3)}({\bf r},{\bf r}') &\equiv& \frac{1}{N}\sum_{i \ne j \ne l}
\langle \delta({\bf r}-{\bf r}_{ij})\delta({\bf r}'-{\bf r}_{il}) \rangle .
\label{eq:g3-def}
\end{eqnarray}

One finds the following continuity equations for the sheared system
relating the partial time derivative of $n_{{\bf q}}(t) = e^{i {\cal L} t} n_{{\bf q}}$
to $j_{{\bf q}}^{\lambda}(t) = e^{i {\cal L} t} j_{{\bf q}}^{\lambda}$: 
\begin{eqnarray}
\Bigl[ 
  \frac{\partial}{\partial t} - 
  {\bf q} \cdot \mbox{\boldmath $\kappa$} \cdot \frac{\partial}{\partial {\bf q}} 
\Bigr]
n_{{\bf q}}(t) = i{{\bf q}}\cdot {{\bf j}}_{{\bf q}}(t),
\label{eq:BK-dum-12}
\end{eqnarray}
and
\begin{eqnarray}
\label{Markov1}
\Bigl[ 
  \frac{\partial}{\partial t} - 
  {{\bf q}} \cdot \mbox{\boldmath $\kappa$} \cdot \frac{\partial}{\partial {\bf q}} 
\Bigr]
j_{{\bf q}}^{\lambda}(t) 
&=&
i q_{\lambda} \frac{v_T^{2}}{S_{q}} n_{{\bf q}}(t) +
i B_{{\bf q}}^{\lambda}n_{{\bf q}}(t) -
A_{{\bf q}}^{\lambda \mu} j_{{\bf q}}^{\mu}(t)
+
R_{{\bf q}}^{\lambda}(t) -
\int_{0}^{t} ds \,
M_{{\bf q}}^{\lambda \mu}(s) \,
e^{i {\cal L} (t-s)} j_{{{\bf q}}(s)}^{\mu} 
\nonumber \\
& & 
+ \,
\int_{0}^{t} ds \,
i L_{{\bf q}}^{\lambda}(s) \,
e^{i {\cal L} (t-s)} n_{{{\bf q}}(s)} -
\int_{0}^{t} ds \,
N_{{\bf q}}^{\lambda \mu}(s) \,
e^{i {\cal L} (t-s)} j_{{{\bf q}}(s)}^{\mu},
\end{eqnarray}
where we have introduced the following memory kernels 
\begin{eqnarray}
M_{{\bf q}}^{\lambda \mu}(t) &\equiv&
\frac{1}{N v_T^{2}}
\langle R_{{\bf q}}^{\lambda}(t) \, R_{{{\bf q}}(t)}^{\mu \, *} \rangle,
\label{eq:M-tr-def}
\\
L_{{\bf q}}^{\lambda}(t) &\equiv&
-i \frac{1}{N S_{q(t)}}
\langle R_{{\bf q}}^{\lambda}(t) \, 
{\cal Q} [ n_{{{\bf q}}(t)}^{*} \Omega(0)] \rangle ,
\label{eq:L-tr-def}
\\
N_{{\bf q}}^{\lambda\mu}(t)&\equiv&
- \frac{1}{N v_T^2}\langle R_{{\bf q}}^{\lambda}(t){\cal Q}[j_{{\bf q}(t)}^{\mu *}\Omega(0)]\rangle .
\label{eq:N-tr-def}
\end{eqnarray}
The equations (\ref{eq:BK-dum-12})-(\ref{eq:N-tr-def}) are the exact equations for uniformly sheared granular liquids
governed by Eqs. (\ref{eq:SLLOD})-(\ref{viscous-force}).

To obtain a closure to these equations,
one has to introduce some approximations such as the mode-coupling approximation.
The details of such approximations will be reported elsewhere.\cite{Hayakawa09}
Instead, here, let us briefly explain what equations can be obtained under the Markovian approximations in the case of
a weak shear and an elastic limit. 

It is straightforward to show that
Eq.(\ref{Markov1}) reduces to 
\begin{equation}
\label{Markov2}
\Bigl[ 
  \frac{\partial}{\partial t} - 
  {{\bf q}} \cdot \mbox{\boldmath $\kappa$} \cdot \frac{\partial}{\partial {\bf q}} 
\Bigr]
j_{{\bf q}}^{\lambda}(t) 
\approx
i q_{\lambda} \frac{v_T^{2}}{S_{q}} n_{{\bf q}}(t)  -
\int_{0}^{t} ds \,
M_{{\bf q}}^{\lambda \mu}(s) \,
e^{i {\cal L} (t-s)} j_{{{\bf q}}(s)}^{\mu} 
+
R_{{\bf q}}^{\lambda}(t)
\end{equation}   
under the weak shear and the elastic limit.
Thus, the effect of shear appears only through the convective deformation of the wave number in the elastic and the unsheared limit.
This equation still includes the non-Markovian memory kernel. 

For many situations in a liquid state far from the jamming transition, one can ignore memory effects.
When we adopt the Markovian approximation with the assumption that the system is isotropic,
it is known that the memory kernel can be approximately given by
its hydrodynamic limit\cite{balucani-zoppi}
\begin{equation}
M_{\bf q}^{\lambda \mu}(t) \approx
\hat{q}_{\lambda} \hat{q}_{\mu} \,  q^{2} \nu_{1} \delta(t) +
(\delta_{\lambda \mu} - \hat{q}_{\lambda} \hat{q}_{\mu}) \, 
 q^{2} \nu_{2} \delta(t) ,
\label{eq:M-Markov-decomposition}
\end{equation}
where $\hat{q}_{\lambda}\equiv q_{\lambda}/q$, $\nu_1$ and $\nu_2$ are the bulk kinetic viscosity and the kinetic viscosity, respectively. 
Equation (\ref{Markov2}) then reduces to 
\begin{equation}
\label{Markov3}
\Bigl[ 
  \frac{\partial}{\partial t} - 
  {{\bf q}} \cdot \mbox{\boldmath $\kappa$} \cdot \frac{\partial}{\partial {\bf q}} 
\Bigr]
j_{{\bf q}}^{\lambda}(t) 
\approx 
i q_{\lambda} \frac{v_T^{2}}{S_{q}} n_{{\bf q}}(t)  -
\nu_1 q_{\lambda} ({\bf q}\cdot{\bf j}_{{\bf q}}(t))-
\nu_2 q^2(\delta_{\lambda\mu}
-\hat{q}_{\lambda}
\hat{q}_{\mu})
 j_{{\bf q}}^{\lambda}(t)
+R_{{\bf q}}^{\lambda}(t) ,
\end{equation}
where $R_{{\bf q}}^{\lambda}(t)$ satisfies the fluctuation-dissipation relation
\begin{equation}
\label{FDR-noise}
\langle R_{{\bf q}}^{\lambda}(t)R_{{\bf k}}^{\mu}(0)\rangle
=\delta(t)\delta({\bf q}+{\bf k})N v_T^2
\{\hat{q}_{\lambda} \hat{q}_{\mu} \,  q^{2} \nu_{1}  +
(\delta_{\lambda \mu} - \hat{q}_{\lambda} \hat{q}_{\mu}) \, 
 q^{2} \nu_{2} \} .
 \end{equation}
Equation (\ref{Markov3}) combined with Eqs. (\ref{eq:BK-dum-12}) and (\ref{FDR-noise}) is the equation of
fluctuating hydrodynamics.

Let us notice that
both the equal-time long-range correlation function\cite{otsuki09a} and the long-time tails of 
autocorrelation functions\cite{otsuki07,otsuki09b} can be discussed within the framework of the
fluctuating hydrodynamics.
Our liquid theory presented here, therefore, provides
not only a microscopic basis of the fluctuating hydrodynamics, but also a basis of both the long-time tails  and 
the long-range correlations for sheared granular liquids. 
 
\section{Discussion and conclusion}

\subsection{Discussion}

Now, let us compare our formulation with previous formulations of the Green-Kubo formula
for granular fluids.\cite{dufty08,vanNoije,dufty02a,dufty02b,lutsko02,brey05}
We first stress that our formulation is unique in that it applies to a nonequilibrium steady state of uniformly sheared granular liquids,
while the previous ones deal with the Green-Kubo formula for granular gases in a freely cooling state.
We also note that Green-Kubo formula in the previous studies is the linear response theory to
a nonequilibrium ``steady state'' (which includes a homogeneous cooling state), but our method is a nonlinear response theory
to the initial canonical state. In practice, the application of the linear response theory to freely cooling granular gases
 has some technical problems; (i) one cannot take the $t\to \infty$ limit of the integral of the time-correlation function
because the granular particles quickly lose their kinetic energy, (ii) thus the behavior of the steady state strongly depends on
the cut-off time of the integration, and (iii) the determination of a nonequilibrium steady distribution is difficult. Let us also notice that
that freely cooling states cannot be realized in experiments.
On the other hand, our method which is the nonlinear response theory to the initial canonical distribution has several advantages
such as (i) the steady state under a uniform shear can be approximately realized in many situations, and (ii) the steady state is
almost independent of the choice of the initial condition, although our method cannot be generalized to a response theory to
a reference state.  
    
Nevertheless, it is remarkable that the Green-Kubo formula for freely cooling granular gases has a similar structure to our generalized Green-Kubo formula.
Indeed, Eq. (139) in ref.\cite{dufty08}  is the essentially same as Eq.(\ref{eq:ss-average}) where their expression
corresponds to the integral 
$\eta\propto \int_0^{\infty} dt \langle \sigma_{xy}(0) [e^{-i {\cal L}^{\dagger}t} \Omega(\Gamma)] \rangle$ in our context. 

\subsection{Conclusion}

This paper summarizes our recent studies on the granular liquid theory under the uniform shear.
We demonstrated that there exists the generalized Green-Kubo formula in sheared granular liquids. 
We also showed that it yields the nonequilibrium steady-state average which is essentially independent
of a specific choice of the initial condition. 
It should be noted that our formulation does not rely on the presence of the steady distribution function which is hard to obtain, but 
any averaged quantity in the steady state can be obtained under the simple initial condition.
In the previous section, we outlined how we can apply our formulation to characterize the behaviors of granular liquids.
One of the most important conclusions in this paper is that the granular liquid theory starting from the Liouville equation reduces
to the fluctuating hydrodynamics 
if one considers a weak-shear and elastic limit along with the Markovian approximation.
Thus, this paper provides a microscopic support of the fluctuating hydrodynamics which is known to
give accurate results of the time correlations and the spatial correlations.

We have skipped detailed derivation of Mori-type generalized Langevin equation and MCT.
This will be discussed in another paper.\cite{Hayakawa09}  
We also note that we can derive the integral fluctuation theorem 
along the same line, though there is no microscopic time reversal symmetry.~\cite{chong09}  
The formulation of granular liquid theory, thus, gives an interesting
subject in pure nonequilibrium statistical mechanics.\cite{chong09b}

\begin{theacknowledgments}
This work was partially supported by Ministry of Education, Culture, Science and Technology (MEXT), Japan
(Nos. 20740245, 21015016, 21540384 and 21540388), and by the Global COE program " The Next Generation of Physics, Spun from
Universality and Emergence" from MEXT Japan. The author also thanks the Yukawa International Program for
Quark-Hadron Sciences at Yukawa Institute for Theoretical Physics, Kyoto University.
The numerical calculation was carried out on Altix3700 BX2 at YITP in Kyoto University.

\end{theacknowledgments}



\bibliographystyle{aipproc}   


\IfFileExists{\jobname.bbl}{}
 {\typeout{}
  \typeout{******************************************}
  \typeout{** Please run "bibtex \jobname" to optain}
  \typeout{** the bibliography and then re-run LaTeX}
  \typeout{** twice to fix the references!}
  \typeout{******************************************}
  \typeout{}
 }

\end{document}